\documentclass[useAMS,usenatbib]{mn2e}

\usepackage{amssymb}
\usepackage{graphicx}

\newcommand{\vc}[1]{\mbox{\boldmath{$#1$}}}

\newcommand{\dpa}{\partial}

\newcommand{\nab}{\vc{\nabla}}

\DeclareMathSymbol{\varOmega}{\mathord}{letters}{"0A}
\DeclareMathSymbol{\varSigma}{\mathord}{letters}{"06}
\DeclareMathSymbol{\varPsi}{\mathord}{letters}{"09}
\DeclareMathSymbol{\varPhi}{\mathord}{letters}{"08}
\DeclareMathSymbol{\varLambda}{\mathord}{letters}{"03}

\newcommand{\Eq}[1]{equation (\ref{#1})}

\newcommand{\Fig}[1]{Fig.~\ref{#1}}

\newcommand{\apj}{ApJ}
\newcommand{\apjl}{ApJ}
\newcommand{\mnras}{MNRAS}
\newcommand{\aap}{A\&A}
\newcommand{\aj}{AJ}
\newcommand{\nat}{Nature}

\newcommand{\apss}{Ap\&SS}

\title[Prograde rotation of protoplanets]{Prograde rotation of protoplanets by
accretion of pebbles in a gaseous environment}

\author[Anders Johansen and Pedro Lacerda]{Anders Johansen$^{1}$\thanks{E-mail:
ajohan@strw.leidenuniv.nl} and Pedro Lacerda$^{2,3}$\\ $^{1}$Leiden Observatory,
Leiden University, P.O.\ Box 9513, 2300 RA Leiden, The Netherlands\\
$^{2}$Queen's University, Belfast, County Antrim, BT7 1NN, United Kingdom\\
$^{3}$Newton Fellow}

\begin{document}

\date{Accepted 200- January --. Received 200- January --; in original form
200- January --}

\pagerange{\pageref{firstpage}--\pageref{lastpage}} \pubyear{2010}

\maketitle

\label{firstpage}

\begin{abstract}
We perform hydrodynamical simulations of the accretion of pebbles and rocks
onto protoplanets of a few hundred kilometres in radius, including two-way drag
force coupling between particles and the protoplanetary disc gas. Particle
streams interacting with the gas far out within the Hill sphere of the
protoplanet spiral into a prograde circumplanetary disc. Material is accreted
onto the protoplanet due to stirring by the turbulent surroundings. We
speculate that the trend for prograde rotation among the largest asteroids is
primordial and that protoplanets accreted 10\%--50\% of their mass from pebbles
and rocks during the gaseous solar nebula phase. Our model also offers a
possible explanation for the narrow range of spin periods observed among the
largest bodies in the asteroid and trans-Neptunian belts, and predicts that
1000 km-scale Kuiper belt objects that have not experienced giant impacts
should preferentially spin in the prograde direction.
\end{abstract}


\section{Introduction}

Planets form in protoplanetary gas discs as dust grains collide and grow to
ever larger bodies \citep{Safronov1969,Dominik+etal2007}. Accumulation of
differentiated meteorite parent bodies must occur while $^{26}$Al and $^{60}$Fe
are still present in the solar nebula, constraining their formation time to at
most a million years after the formation of calcium-aluminium-rich inclusions
(\citealt{Bizzarro+etal2005}; \citealt*{Yang+etal2007}). This coincides with
the gaseous solar nebula epoch and indicates that planetesimals and
protoplanets of several hundred kilometres in size were accumulated in a dense
gaseous environment.

The alignment between the orbits of the planets and the rotation of the Sun is
one of the key pieces of evidence for the nebular hypothesis of the formation
of the solar system. The large majority of satellites also orbit their host
planet in what is termed the prograde direction, i.e.\ orbit in the same
direction as the planet orbits the sun, suggesting that they formed in
circumplanetary discs. This regularity seems to extend to the predominantly
prograde rotations of the planets and the largest asteroids. While the gas
giant planets have likely been spun up by gas accretion, growth of solid
planetary bodies by accretion of km-sized planetesimals is not generally
expected to lead to systematic prograde rotation. Post-accretion, giant impacts
which can significantly change the obliquity and spin rate of protoplanets
\citep{DonesTremaine1993b,KokuboIda2007} do so in a stochastic way yielding no
preference for prograde motion. Successful attempts to explain a {\it
systematic} prograde contribution from accretion of solids in a gas-free medium
require special surface density profiles \citep{LissauerKary1991} or
``semicollisional'' accretion \citep{SchlichtingSari2007}, i.e. collisions
between small particles and subsequent accretion into a prograde
circumplanetary particle disc that feeds the central object.
\begin{figure*}
  \begin{center}
    \includegraphics[width=12.0cm]{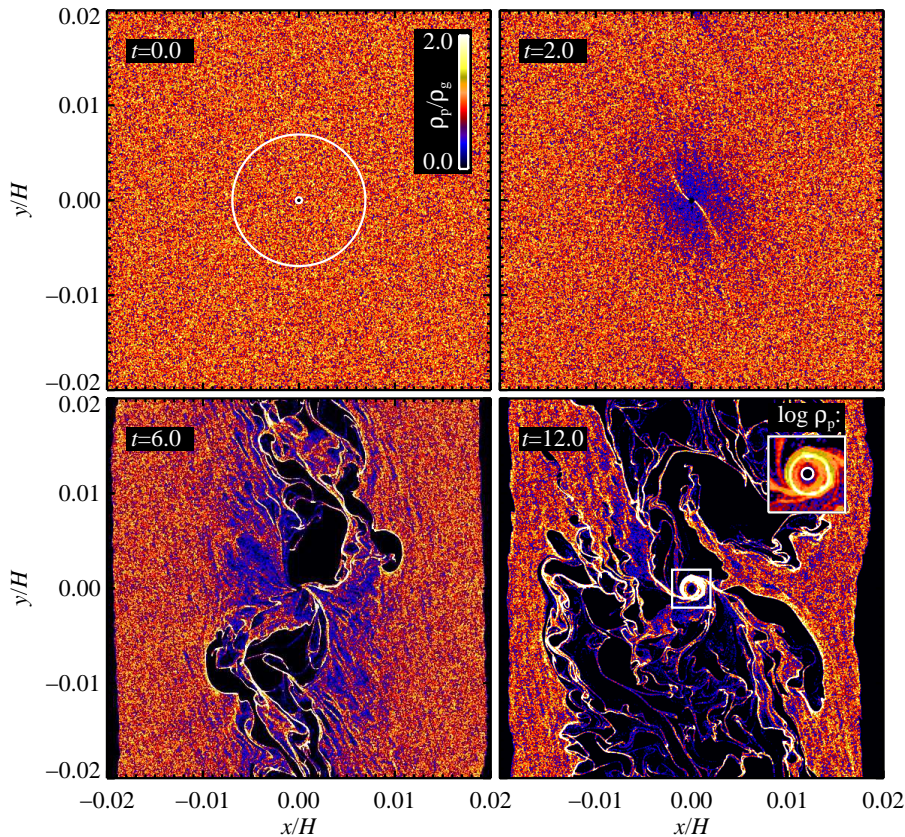}
  \end{center}
  \caption{Accretion of pebbles onto a protoplanet fixed in the centre of the
  coordinate system. The first panel indicates the Hill radius and the
  accretion radius (inner hole). The colour coding shows the local density of
  solids, in units of the mid-plane gas density. Particles are initially
  accreted from the second and fourth quadrants (second panel). As pebble
  streams drag the gas down towards the protoplanet, the incompressible gas
  expands in great particle-free bubbles, eventually making the system
  turbulent (third panel). A prograde accretion disc, fed from the second and
  fourth quadrants, forms around the protoplanet (fourth panel).}
  \label{f:mainresult}
\end{figure*}
\begin{figure*}
  \begin{center}
    \includegraphics[width=0.7\linewidth]{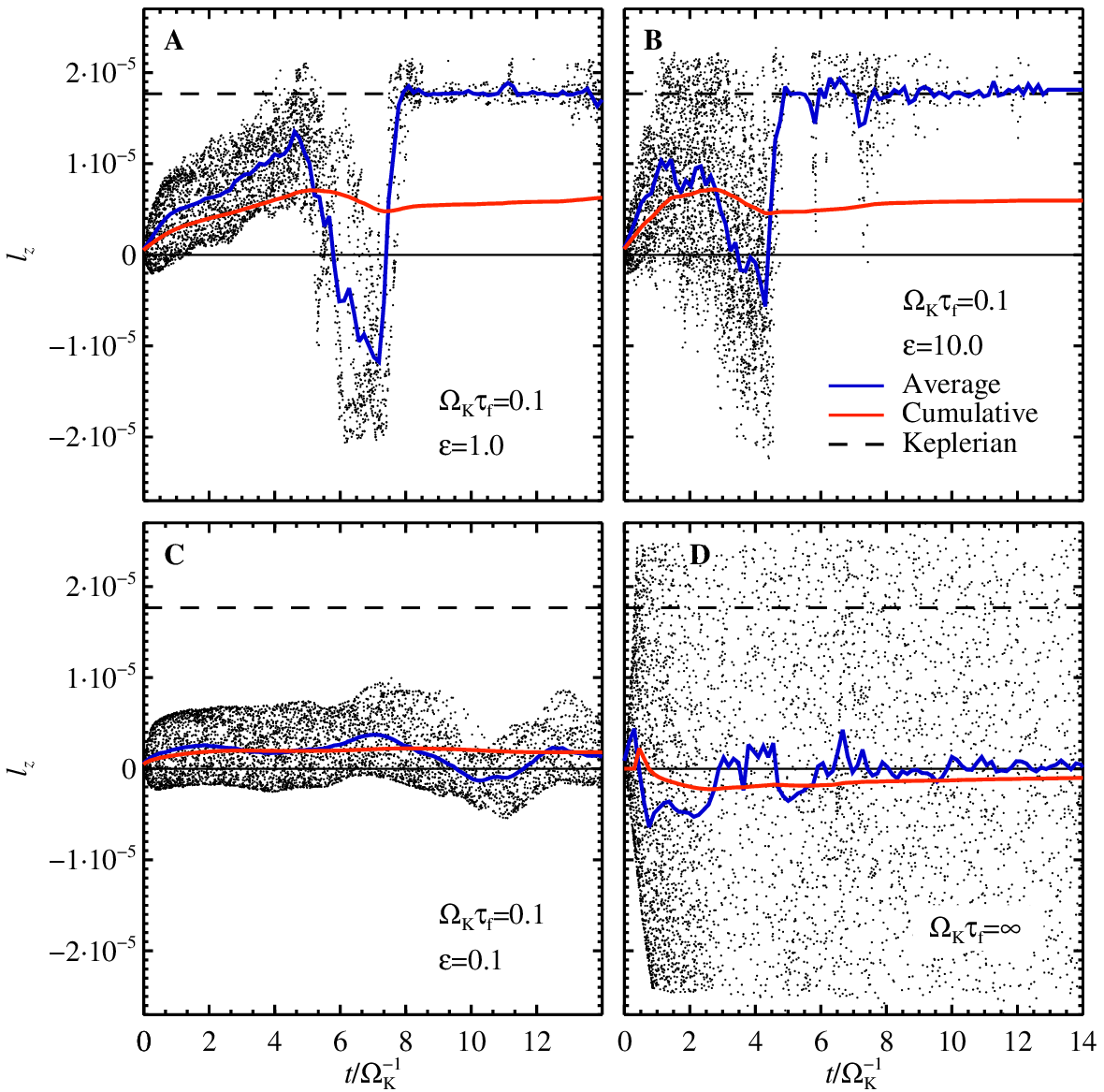}
  \end{center}
  \caption[]{Delivery of prograde angular momentum to a protoplanet of 400 km
  radius (if placed at $r_0=10\,{\rm AU}$). The dots show the angular momentum
  of a subset of the accreted particles, while the blue (black) and red (grey)
  lines indicate the instantaneous and cumulative average. ({\bf A}-{\bf B})
  Accretion of pebbles from environment with mean solids-to-gas ratio
  $\epsilon=1.0$ and $\epsilon=10.0$, respectively. The presence of a Keplerian
  particle disc is clearly visible at late times where the accreted angular
  momentum is exactly equal to that of a Keplerian disc (dashed line). ({\bf
  C}) In the low solids-to-gas ratio case ($\epsilon=0.1$) the pebbles have
  little influence on the gas and their motion is strongly damped as they
  approach the protoplanet, imparting only little rotation to the protoplanet.
  ({\bf D}) Without drag forces these effectively very large particles give a
  net retrograde rotation to the protoplanet.}
  \label{f:angmom_deliv}
\end{figure*}

In this paper we show that prograde rotation may be a natural consequence of
planetesimal growth from accretion of cm-sized solids in a gaseous
environment. Providing a physical mechanism for prograde rotation in turn
supports that the dominance of prograde rotation among the largest asteroids is
significant despite their low number. Our model assumes simultaneous presence
of cm-sized ``pebbles'' and gas. Such conditions are supported by observations
of protoplanetary discs, which show evidence of dust growth to at least a few
cm \citep{Testi+etal2003,Sicilia-Aguilar+etal2007}. The inferred population of
mm-cm pebbles can be very substantial, on the order of $10^{-3}$ solar masses
\citep{Wilner+etal2005}.

We present hydrodynamical computer simulations of the accretion of pebbles and
rocks onto young protoplanets in a gaseous nebula environment. We find that
particle streams interacting with the gas far out within the Hill sphere
dissipate enough energy that the particles spiral down towards the protoplanet
and form a prograde accretion disc. The prograde motion is transferred to
protoplanet rotation as the particles are accreted. To explain the observed
rotations of the largest asteroids and Kuiper belt objects, we propose that
protoplanets accreted a major fraction (10\%--50\%) of their mass from pebbles
and rocks during the gaseous solar nebula phase.

The paper is organised as follows. In \S\ref{s:setup} we describe the set up of
our simulations. In \S\ref{s:results} the main results are presented, both for
2-D and 3-D simulations. We speculate in \S\ref{s:discussion} about
implications for the observed prograde rotation of large asteroids and the
future prospect of measuring rotation directions of Kuiper belt objects. We
conclude on our findings in \S\ref{s:conclusions}. The appendices contain
details that do not appear in the main paper and several tests of the validity
of our results. In Appendix \ref{s:dyneq} the dynamical equations are
presented. We validate the code in Appendix \ref{s:validate} against known
results on the accreted angular momentum. Appendix \ref{s:mmsn} describes drag
forces in more detail and also the translation from dimensionless friction time
to physical particle size. In Appendix \ref{s:collide} we compare time-scales
of drag force and particle collisions and conclude that drag forces are by far
the dominant force, except in regions that are highly particle dominated. In
Appendix \ref{s:converge} we present results at lower and higher resolution and
conclude that prograde rotation is strengthened with increasing resolution. The
dependence of the results on the size of the simulation box are treated in
Appendix \ref{s:boxsize}.

\section{Simulation set up}
\label{s:setup}

We perform numerical simulations of the coupled motion of solids and gas in a
local coordinate frame corotating with the disc at the Keplerian frequency
$\varOmega_{\rm K}$ at an arbitrary distance $r_0$ from the central star
\citep{GoldreichTremaine1980}. Coordinate axes are oriented such that $x$
points radially away from the star and $y$ points along the rotation direction
of the disc. Particles are treated by a Lagrangian method, interacting with the
gas through two-way drag forces operating on the friction time-scale $\tau_{\rm
f}$. An additional gravitational force contribution from a protoplanet of mass
$M_{\rm p}$ fixed in the centre of the coordinate frame is added. The gas is
evolved on a fixed grid and also feels the gravity of the protoplanet (see
Appendix \ref{s:dyneq} for details on the dynamical equations). In Appendix
\ref{s:validate} we validate the results of the code against known results on
angular momentum accretion \citep{SchlichtingSari2007}.

Momentum conservation in the two-way drag force coupling between solid
particles and gas is assured by assigning the drag force added on a particle
back to the gas with the opposite sign by a second order scheme. This drag
force scheme has been shown to successfully reproduce the linear growth rate of
streaming instabilities in protoplanetary discs \citep{YoudinJohansen2007}.

We integrate the dynamical equations of particles and gas using the high order
finite difference code Pencil Code \citep{Brandenburg2003}. To avoid
underresolution of motion near the protoplanet, we always freeze the gas inside
a ring of diametre eight grid points around the protoplanet. Particles crossing
that ring are considered to be accreted onto the protoplanet.

The Hill radius, $R_{\rm H}=(GM_{\rm p}/3\varOmega_{\rm K}^2)^{1/3}$ where $G$
is the gravity constant, bounds the region where the gravity of the protoplanet
dominates over the tidal force from the star. In a protoplanetary disc with
vertical pressure scale height $H$, a protoplanet with mass $G M_{\rm
p}=10^{-6}\varOmega_{\rm K}^2H^3$ has $R_{\rm H}\approx0.007 H$. At
$r_0=10\,{\rm AU}$, with disc aspect ratio $H/r_0=0.06$, the protoplanet has
mass $M_{\rm p}\approx4\times10^{23}\,{\rm g}$ and radius $R_{\rm
p}\approx400\,$km. At $r=3\,{\rm AU}$ (disc aspect ratio $H/r_0=0.043$) the
protoplanet mass would be $M_{\rm p}\approx1.6\times10^{23}\,{\rm g}$ and its
radius $R_{\rm p}\approx290\,$km.

We use a standard box size of $(L_x,L_y)=(0.04H,0.04H)$, at a standard
resolution of $512^2$ grid points and $10^6$ particles. This gives a resolution
element of around 5,000 kilometres, making the inner cavity 20,000 kilometres
in radius at the standard resolution. In Appendix \ref{s:converge} we present
$256^2$ and $1024^2$ simulations. The inner cavity corresponds to 1/10, 1/20
and 1/40 of the Hill radius for low, medium and high resolution studies,
respectively. The convergence test in Appendix \ref{s:converge} shows that the
prograde rotation is strengthened at higher resolution with a correspondingly
smaller inner accretion radius.

\section{Results}
\label{s:results}

In Figure \ref{f:mainresult} we show results of a simulation with an assumed
solids-to-gas ratio $\epsilon$ of unity (mimicking conditions in a sedimentary
mid-plane layer) and a friction time $\varOmega_{\rm K}\tau_{\rm f}=0.1$
(relative to the inverse Keplerian orbital frequency). This friction time
represents particles of a few cm in radius \citep{Weidenschilling1977}. See
Appendix \ref{s:mmsn} for details. The total mass of particles in the
simulation domain is approximately $1.8\times10^{22}\,{\rm g}$ at $r=3\,{\rm
AU}$ (with particle column density $\varSigma_{\rm p}=3\,{\rm g\,cm^{-2}}$) and
$6.4\times10^{22}\,{\rm g}$ at $r=10\,{\rm AU}$ (with $\varSigma_{\rm
p}=0.5\,{\rm g\,cm^{-2}}$).

As time progresses a spiral density wave structure appears inside the Hill
sphere. Particles are accreted primarily from the second and fourth quadrants.
As the particles drag the gas along towards the protoplanet surface, and the
gas is nearly incompressible, the gas leaves through a ``wind'' in the first
and third quadrants. This wind prevents particles from entering from these
sides.

Later in the simulation the particle-gas layer becomes turbulent, due to the
bubbles of gas expanding from the protoplanet surface. A prograde particle disc
forms around the protoplanet, still fed from the second and fourth quadrants.
This disc is completely dominated by particles ($\rho_{\rm p}/\rho_{\rm
g}\gg10$) and accretes slowly, due to stirring by the turbulent surroundings.
We show in Appendix \ref{s:collide} that while particle collisions are
negligible in most of the simulation domain, collisions will dominate over gas
drag in the circumplanetary disc. However, we do not model the collisional
viscosity in the presented simulations, as we wish to keep the system as simple
as possible. A numerical algorithm for treating collisions between
superparticles is currently under development (Johansen, Lithwick, \& Youdin,
in preparation).

In Figure \ref{f:angmom_deliv} we plot the angular momentum of accreted
particles as a function of time, for a subset of randomly sampled particles.
Initially, during the spiral density wave phase, there is a preference for
accreting prograde particles. This is normally followed by a short phase of
retrograde accretion, where a retrograde accretion disc may temporarily form.
Eventually the particles always enter a prograde accretion disc, and the
protoplanet accretes the full Keplerian surface frequency.

Convergence tests are presented in Appendix \ref{s:converge}. Both lower and
higher resolution studies result in a similar amount of prograde accretion,
although there is a trend for accreting stronger prograde rotation with
increasing resolution. In Appendix \ref{s:boxsize} we present results for a box
size of $0.1H\times0.1H$. The larger box size has a net positive effect on the
prograde rotation of the protoplanet. The sign of the angular momentum is more
consistently positive than in the small box, and there are fewer periods of
retrogade accretion.

\subsection{Dependence on particle size}

\begin{figure}
  \begin{center}
    \includegraphics[width=7.0cm]{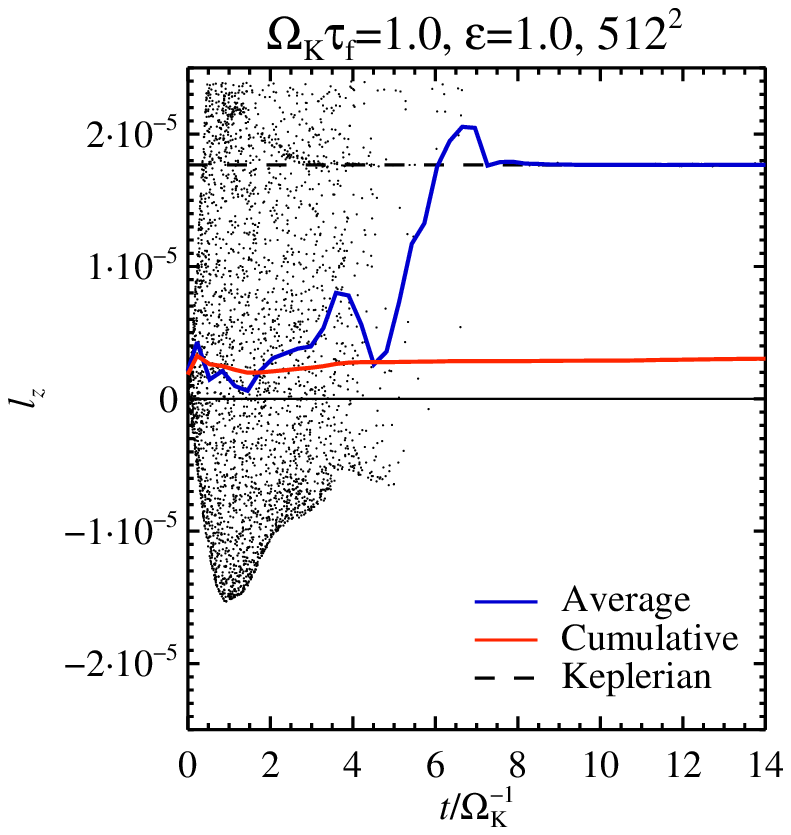}
    \includegraphics[width=7.0cm]{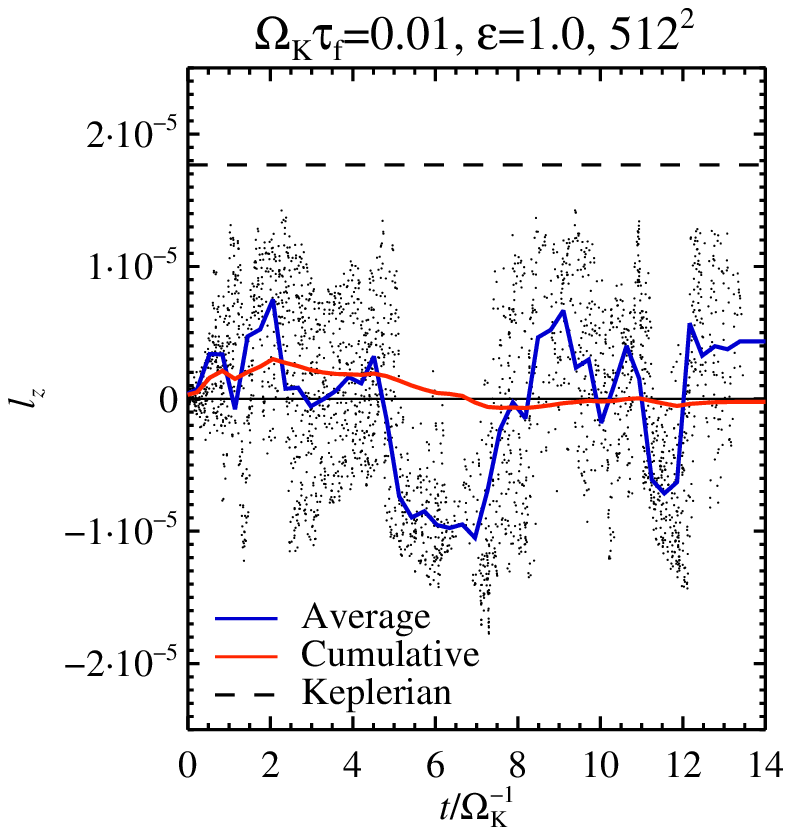}
  \end{center}
  \caption{Accretion of larger and smaller particles. Marginally coupled
  particles $\varOmega_{\rm K}\tau_{\rm f}=1$ particles (top plot) readily form
  a prograde particle disc. The particles are much less affected by the gas
  than the particles presented in \Fig{f:angmom_deliv}, thus the particle disc
  accretes very slowly. Smaller particles with $\varOmega_{\rm K}\tau_{\rm
  f}=0.01$, corresponding to sizes of a few mm, are so coupled to the gas that
  accretion takes place as a sedimentation towards the protoplanet, with a
  significant loss of angular momentum to the gas (bottom plot).}
  \label{f:lz_t_size}
\end{figure}

It was shown analytically by \cite{MutoInutsuka2008} that particles approaching
the protoplanet feel the opposing effects of protoplanet gravity (at large
distances) and gravitational scattering (at short distances). For particles
with $\varOmega_{\rm K}\tau_{\rm f}<1$, drag forces are so strong that
particles are accreted into the Hill sphere, while larger particles are
scattered by the planet and only approach the protoplanet to some minimum
distance. Similarly \cite{WeidenschillingDavis1985} showed that planetesimals
above a certain size are trapped in mean motion resonances with growing
planets. This shepherding might affect boulders of sizes metres to ten metres.

In \Fig{f:lz_t_size} we show the accretion of larger, $\varOmega_{\rm
K}\tau_{\rm f}=1$, and smaller, $\varOmega_{\rm K}\tau_{\rm f}=0.01$,
particles, corresponding to a few metres and a few millimetres, respectively.
The larger particles readily form a prograde particle disc around the
protoplanet. However, the marginal coupling to the gas makes the particle disc
much more stable to dynamical shaking by the surrounding gas, and the disc only
accretes slowly. Smaller particles, on the other hand, are very coupled to the
gas, and accretion takes place as a slow sedimentation. The particles lose a
significant fraction of their angular momentum to the gas and arrive almost
perpendicular to the protoplanet surface.

Thus we can put a relatively tight constraint on the particle sizes that can
lead to prograde rotation of protoplanets. A dimensionless friction time
between approximately $0.1$ and $1.0$ corresponds to particles of radii between
a few cm and a few ten cm. However, at several ten AU from the central star,
where the gas column density drops to $\varSigma_{\rm g}\approx 5\,{\rm
g\,cm^{-2}}$, the ``good'' particle sizes are between a few mm and a few cm,
approximately corresponding to the pebbles that are observed in the outer
regions of protoplanetary discs
\citep{Testi+etal2003,Wilner+etal2005,Rodmann+etal2006,Lommen+etal2009}.

\subsection{Sub-Keplerian motion}
\label{s:sandblast}

\begin{figure}
  \begin{center}
    \includegraphics[width=7.0cm]{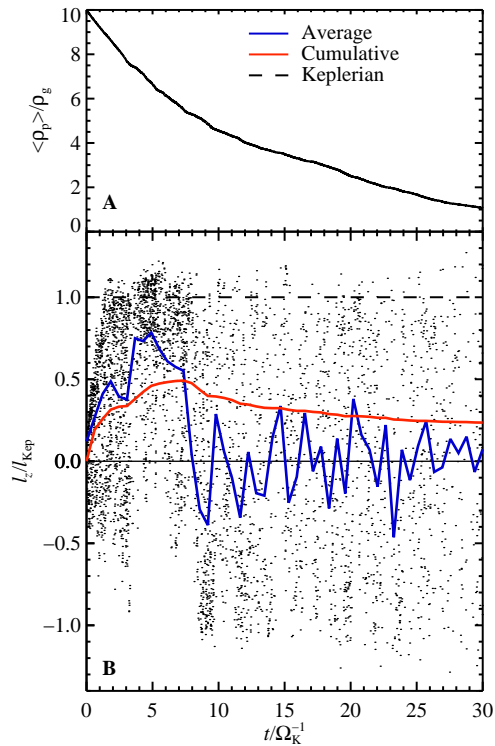}
  \end{center}
  \caption{Sandblasting a protoplanet with pebbles. ({\bf A}) Mean particle
  density in the box as a function of time, ({\bf B}) the accreted angular
  momentum. At the initial solids-to-gas ratio of 10, a net positive angular
  momentum is accreted by the protoplanet. But as the particles are accreted,
  the remaining particles drift stronger and stronger relative to the
  protoplanet, eventually leading to ballistic accretion with no net angular
  momentum.}
  \label{f:sandblast}
\end{figure}

An additional effect that we have not considered so far is the sub-Keplerian
motion of the gas. As the gas is slightly pressure supported in the radial
direction, the gas orbital motion is slower (by approximately 5\% of the sound
speed or 0.5\% of the Keplerian speed) than the Keplerian value
(\citealt*{Adachi+etal1976}; \citealt{Weidenschilling1977}). Isolated particles
face a constant headwind of the gas, causing them to orbit slower and to drift
radially. The stationary protoplanet is thus exposed to a ``sandblast'' with
velocity $\sim 25\,\rm{m}\,\rm{s}^{-1}$. In \Fig{f:sandblast} we consider a
simulation of $\varOmega_{\rm K}\tau_{\rm f}=0.1$ particles with an initial
solids-to-gas ratio of 10. The strong mass loading of particles reduces the
head wind significantly in the beginning, and we observe the usual prograde
accretion. However, as the particles lose influence on the gas, a sandblasting
phase occurs in which no net rotation is induced on the protoplanet. Overall
the protoplanet nevertheless accretes a significant fraction of the Keplerian
surface frequency.

\subsection{3-D results}

\begin{figure}
  \begin{center}
    \includegraphics[width=7.0cm]{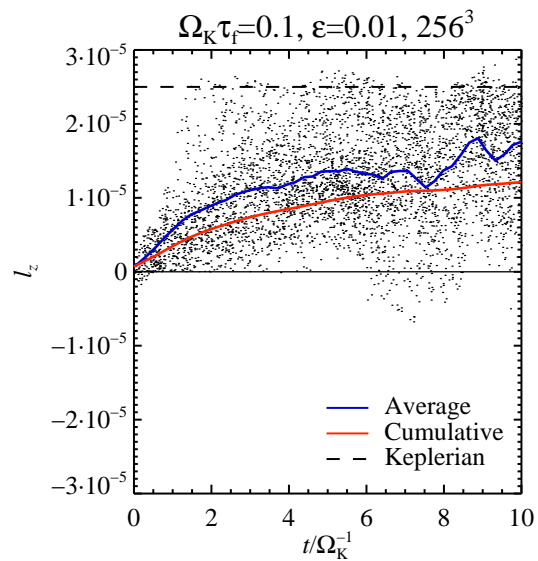}
  \end{center}
  \caption{3-D accretion. The 3-D simulation has resolution $N_x=N_y=N_z=256$
  and $N_{\rm p}=8\times10^6$. We recover the usual preference for prograde
  accretion, showing that the higher resolution 2-D simulations presented in
  the \Fig{f:angmom_deliv} represent the physical system well.}
  \label{f:3Dacc}
\end{figure}
\begin{figure*}
  \begin{center}
    \includegraphics[width=0.7\linewidth]{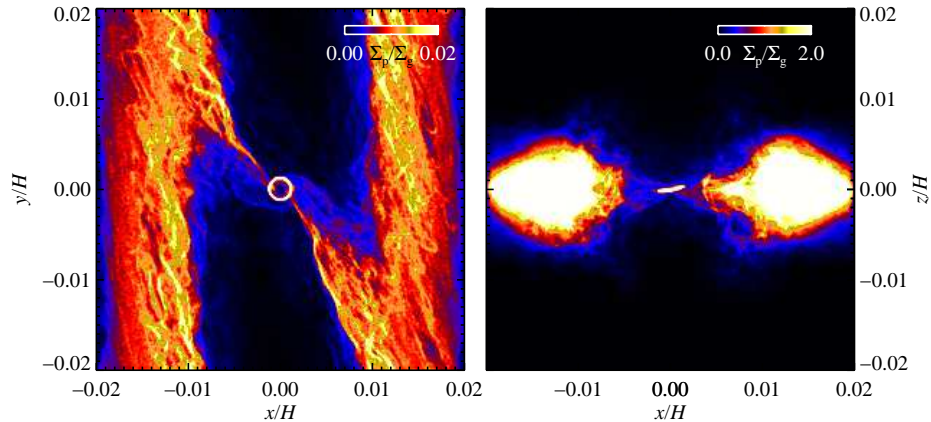}
  \end{center}
  \caption{3-D structure. The figure shows the column density of particles in
  the $x$-$y$ plane (left panel) and in the $x$-$z$ plane (right panel). The
  column density has been normalized by the gas column density in the relevant
  direction. A thin particle disc orbits close to the protoplanet, surrounded
  by the thicker particle mid-plane layer outside the Hill sphere.}
  \label{f:3D}
\end{figure*}

For simplicity, and to obtain higher resolution and smaller accretion radius,
we have so far presented results of 2-D simulations. In this section we show
that we obtain similar results in 3-D as in 2-D. We ignore the effect of
vertical gravity and stratification on the gas, a small effect in boxes with a
vertical extent of $L_z/2=0.02 H$. We initialize the particles with a Gaussian
density distribution around the mid-plane, of scale height $H_{\rm p}=0.01 H$,
and a column density of $0.01$ relative to the gas. The results of a 3-D
simulation with $256^3$ grid points and $8\times10^6$ particles are shown in
\Fig{f:3Dacc}. The particles show the usual strong preference for prograde
accretion. A prograde accretion disc also forms around the protoplanet in the
3-D simulation (see \Fig{f:3D}), but this is not as evident from the accreted
angular momentum as in 2-D because isolated particles, now entering from all
directions, are accreted faster than the disc material. We conclude that our
assumption of 2-D flow represents the true 3-D physical behaviour of the system
well. The 2-D assumption, at the same time, allows for much higher resolution
investigations and a higher particle number per grid cell.

\section{Discussion}
\label{s:discussion}

\subsection{Prograde rotation in the solar system}

\begin{figure}
  \begin{center}
    \includegraphics[width=5.0cm]{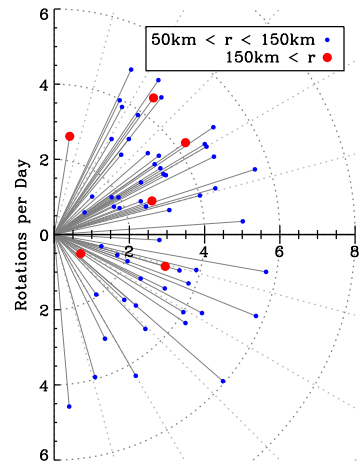}
  \end{center}
  \caption{Obliquity of asteroids larger than 50 km in radius with respect to
  the ecliptic plane ($x$-axis). Each asteroid is represented by a blue (black)
  dot, connected to the origin by a grey line. The angle between each grey line
  and the positive $y$-axis is the obliquity of the corresponding asteroid. The
  length of the lines gives the spin frequency of the asteroid in rotations per
  day. Asteroids in the upper quadrant are prograde and asteroids in the lower
  quadrant are retrograde. Asteroids larger than $r=150$ km, represented by
  larger, red (dark grey) dots, show a preference for prograde rotation.}
  \label{f:poles}
\end{figure}

\begin{figure}
  \begin{center}
    \includegraphics[width=0.8\linewidth]{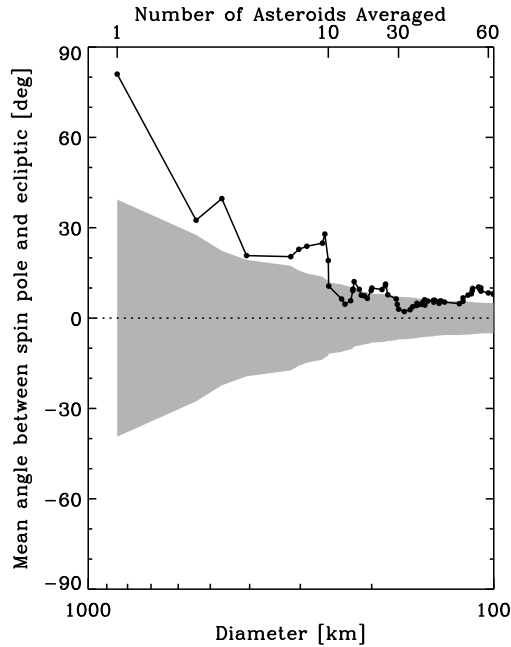}
  \end{center}
  \caption{Mean pole angle of asteroids as a function of their size in
  kilometres. The black dots show the cumulative mean angle from the ecliptic.
  We connect the dots with a line to guide the eye. The first dot from the left
  is the pole angle of Ceres. The second dot is the mean of Ceres and Pallas,
  and so on. The shaded region is the 1-$\sigma$ region expected from random
  spin orientations. The line is above the grey area until approximately the
  10th largest asteroid, although low significance prograde rotation persists
  even for smaller asteroids.}
  \label{f:prograde_significance}
\end{figure}

The asteroid belt is a collisionally evolved collection of rocky and icy bodies
residing roughly between the orbits of Mars and Jupiter. Analytical and
numerical studies indicate that the largest asteroids have not suffered enough
collisional evolution to significantly alter their spin vector
\citep{McAdooBurns1973,Farinella+etal1992}. These objects have thus likely
retained their primordial spin angular momentum and may be representative of
the original protoplanet population \citep[see also][]{Morbidelli+etal2009}.
Figure \ref{f:poles} depicts a collection of spin axis obliquities of 62
asteroids larger than 50 km in diametre \citep[from][]{Kryszczynska+etal2007}.
The obliquity is here defined as the angle between the pole vector of an
asteroid and the normal to the ecliptic plane. While smaller asteroids have
collisionally randomized obliquities, the largest ones ($r>150$ km) display a
propensity towards prograde rotation (obliquity $<90^\circ$), i.e.\ rotation in
the same direction as the orbit. Indeed, the two most massive asteroids, Ceres
and Vesta, spin progradely with obliquities of $9^\circ$ and $36^\circ$ and
periods 9.1 and 5.3 hours. If their spin orientations were collisionally
randomized, the probability that their mean obliquity is less than the observed
value is only $p\sim2.5\times10^{-3}$ (3$\sigma$). In
\Fig{f:prograde_significance} we show the mean pole angle (relative to the
ecliptic) of asteroids as a function of their size. There is a trend towards
prograde rotation for all size ranges, although the statistical significance is
only $1$--$2$ $\sigma$.

\begin{figure}
  \begin{center}
    \includegraphics[width=7.5cm]{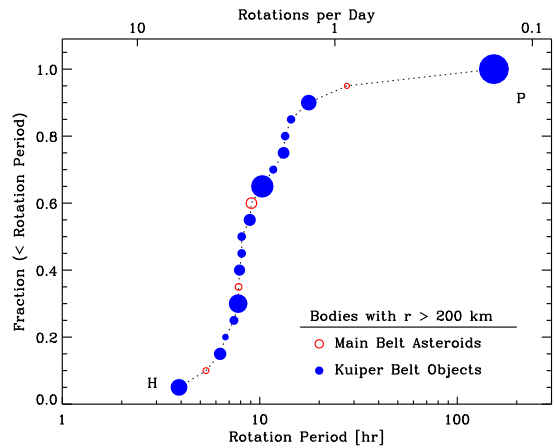}
  \end{center}
  \caption{Cumulative spin period distribution of Kuiper belt objects and
  asteroids larger than 200 km in radius. Most objects lie in a narrow $5<P<20$
  hour period range irrespective of their other physical properties. Outliers
  Haumea (H) and Pluto (P) are marked; they are believed to have suffered
  massive collisions.}
  \label{f:spins}
\end{figure}

The Kuiper belt, a dynamically stable region beyond Neptune, preserves roughly
a dozen known 1000 km-scale icy bodies
\citep{Jewitt1999,Chiang+etal2007,Brown2008}. As in the case of asteroids,
collisional evolution simulations of Kuiper belt objects (KBOs) show that
bodies larger than $r\approx 200$~km have sustained relatively mild collisional
evolution \citep{DavisFarinella1997} and their primordial spins have survived
the random effect of impacts for the age of the solar system
\citep{Lacerda2005}. No KBO obliquities have been derived as the known objects
have moved less than $30^\circ$ along their orbits since discovery, but several
rotation periods have been measured (Fig. \ref{f:spins}). With the exception of
Pluto and Haumea, both believed to have suffered massive collisions
\citep{Canup2005,Brown+etal2007}, the rotation periods of KBOs larger than
$r=200$ km lie in a narrow 5 to 20 hour range (\citealt{LacerdaLuu2006};
\citealt*{Sheppard+etal2008}). The same is true for Ceres and the other three
asteroids in the same size range, which have orbital timescales one to two
orders of magnitude shorter than KBOs. Such uniformity in spin period for
bodies with different compositions and presumably formed and evolved at very
different heliocentric distances suggests a common spin-up mechanism. 

\subsection{Spin-up of protoplanets}

The numerical experiments presented in \S\ref{s:results} show that accretion of
particle sizes of a few cm to a few m and a solids-to-gas ratio around unity or
higher (see Figure 3) leads to a systematic prograde rotation of protoplanets,
in agreement with the observed trend in the largest asteroids. This moderately
high solids-to-gas ratio points to the sedimentary mid-plane, or to dense
particle clumps appearing in such a mid-plane layer
(\citealt{Johansen+etal2007}; \citealt*{Cuzzi+etal2008}), as a prime site for
the growth of protoplanets. The evaporation of gas by irradiation from the
central star or neighbouring O stars leads to a similar increase in the
influence of the dust particles on the gas
\citep*{ThroopBally2005,AlexanderArmitage2007,Johansen+etal2009}.

Our simulations predict that the largest Kuiper belt objects should
preferentially rotate in the prograde direction around their axis. A few
exceptions to this trend are objects that have suffered massive impacts
\citep[e.g. Haumea, ][]{Brown+etal2007}.

Pure accretion from a circumplanetary disc would lead to a protoplanet rotation
period of only a few hours, near the centrifugal break up speed. However,
accretion from a circumplanetary disc need not be the only mass source. Growth
phases with little or no accreted angular momentum can occur when the
protoplanet accretes isolated dust particles and pebbles or km-sized
planetesimals or other protoplanets. Protoplanets thus need to obtain between
10\% and 50\% of their mass from the accretion of macroscopic dust particles in
a gaseous environment. The lower limit is relevant if all pebbles are accreted
at the full Keplerian angular momentum, while the upper limit is relevant if
80\% of the mass accretion occurs through pebble sandblasting or through
accretion of ``planetesimals'' with high $\tau_{\rm f}$.

The sub-Keplerian sandblasting effect described in \S\ref{s:sandblast} means
that the mechanism for obtaining prograde rotation works best in
particle-dominated flows or in regions of low radial drift. The radial drift
can be reduced or even reversed in persistent pressure bumps appearing in the
gas \citep{HaghighipourBoss2003}. Recent studies have shown how planetesimal
formation by self-gravity \citep{Johansen+etal2007,Kato+etal2009} and by
coagulation (\citealt{KretkeLin2007}; \citealt*{Brauer+etal2008}) proceed in
such convergence regions in the flow of solid particles.

\subsection{Spin-up in absence of gas?}

It was recently proposed by \cite{SchlichtingSari2007} that inelastic particle
collisions will lead to accretion of prograde rotation from a circumplanetary
particle disc during the gas-free stages of planetary growth. Such a mechanism
may be able to provide additional prograde angular momentum after the giant
impact stage. This is especially relevant for the terrestrial planets which are
believed to have been accumulated over 10-100 million years in a gas free
environment \citep{Lunine+etal2010}. Mercury, Earth and Mars all have prograde
spins, but the spins of the terrestrial planets have been substantially
modified by long-term perturbations by other planets \citep{LaskarRobutel1993}.

Focusing instead on the largest asteroids and Kuiper belt objects, our
numerical simulations show that pebbles and rocks are accreted very efficiently
by the protoplanets. In the main numerical simulation presented in
\S\ref{s:results} the protoplanet+disc system accretes 1/7 of the protoplanet
mass in five orbits ($\approx$ 25 years at 3 AU). Such high accretion rates are
facilitated by the presence of gas in the Hill sphere. This indicates that
young protoplanets can accrete lots of material already during the
protoplanetary gas disc phase. The actual ratio of accretion during and after
the gaseous disc phase is nevertheless relatively unconstrained and deserves
further attention in the future.

\section{Conclusions}
\label{s:conclusions}

We present computer simulations of the accretion of pebbles and rocks onto
young protoplanets in a gaseous nebula environment. Our findings can be
summarised as follows:

\begin{itemize}
  \item Particle streams interacting with the gas far out in the Hill sphere of
  the growing protoplanet dissipate enough energy that the particles spiral
  down towards the central object and form a prograde accretion disc. The
  prograde motion is transferred to the protoplanet as the particles are
  accreted.
  \item In order to explain the observed rotations of the largest asteroids and
  Kuiper belt objects, we propose that protoplanets accreted a major fraction
  (10\%--50\%) of their mass from pebbles and rocks during the gaseous solar
  nebula phase.
  \item A specific prediction of our model is that most of the largest Kuiper
  belt objects should possess prograde spins.
\end{itemize}

Since the Keplerian surface frequency depends only on the density of the
protoplanet, accretion from a Keplerian particle disc explains how asteroids
and Kuiper belt objects can have similar rotation periods even though the
orbital time-scale is much longer in the Kuiper belt. Our results imply not
just a universality in the rotation period of solid bodies in the solar system,
but also in the formation process and in the important role of gas and pebbles
for the growth of hundred kilometre scale protoplanets.

Future numerical work should focus on the extension of the presented model to
include turbulence driven by global instabilities and more realistic solid
physics, such as a distribution of particle sizes, collisional viscosity and
collisional fragmentation.

\section*{Acknowledgments}

Computer simulations were performed at the PIA cluster of the Max Planck
Institute for Astronomy. PL is grateful for the support of a Royal Society
Newton Fellowship. The authors would like to thank Hilke Schlichting for
reading and commenting an early version of the manuscript.

\appendix

\section{Dynamical equations}
\label{s:dyneq}

We simulate the coupled motion of particles and gas in the shearing sheet frame
\citep{GoldreichTremaine1980}. The frame rotates at the Keplerian frequency
$\varOmega_{\rm K}$ at an arbitrary distance $r_0$ from the central star. Axes
are oriented such that $x$ points radially away from the star, $y$ points along
the rotation direction of the disc, while $z$ points vertically out of the
orbital plane. The particle component is treated as Lagrangian particles, each
with a position vector $\vc{x}=(x,y)$ and a velocity vector $\vc{v}=(v_x,v_y)$,
with dynamical equations
\begin{eqnarray}
  \dot{x}  &=& v_x \, , \\
  \dot{y}  &=& v_y - (3/2) \varOmega_{\rm K} x \, , \\
  \dot{v}_x &=& +2\varOmega_{\rm K} v_y-\frac{G M_{\rm p}
  x}{(x^2+y^2)^{3/2}} - \frac{1}{\tau_{\rm f}} (v_x - \overline{u}_x)
  \label{eq:xddot} \, , \\
  \dot{v}_y &=& -\frac{1}{2}\varOmega_{\rm K} v_x-\frac{G M_{\rm p}
  y}{(x^2+y^2)^{3/2}} - \frac{1}{\tau_{\rm f}} (v_y - \overline{u}_y) \label{eq:yddot} \, .
\end{eqnarray}
We measure the velocity in the $y$-direction relative to the linearised
Keplerian shear $u_y^{(0)}=-(3/2)\varOmega_{\rm K} x$, which gives rise to a
change in the Coriolis force appearing in equation~(\ref{eq:yddot}). An
additional force contribution from the protoplanet of mass $M_{\rm p}$ fixed in
the centre of the coordinate frame is added. We also consider the drag force
from the gas, acting on the friction time-scale $\tau_{\rm f}$ and proportional
to the velocity difference between particles and gas. The gas velocity
$\overline{\vc{u}}=(\overline{u}_x,\overline{u}_y)$ at the position of a
particle is found by spline interpolation from the nearest three grid points in
each direction \citep{YoudinJohansen2007}.

The evolution of gas is solved on a fixed Cartesian grid. The equations of
motion for the gas velocity field $\vc{u}$, relative to the Keplerian shear,
and the continuity equation for gas density $\rho_{\rm g}$, read
\begin{eqnarray}
  \frac{\dpa u_x}{\dpa t} + (\vc{u}\cdot\nab) u_x + u_y^{(0)} \frac{\dpa
    u_x}{\dpa y}&=& +2\varOmega_{\rm K} u_y - 2 \varOmega_{\rm K} \Delta
    v \nonumber \\
    && \hspace{-4.0cm} - \frac{G M_{\rm p} x}{(x^2+y^2)^{3/2}} - c_{\rm s}^2
    \frac{\dpa \ln \rho_{\rm g}}{\dpa x} - \frac{\epsilon}{\tau_{\rm f}}
    (u_x-\overline{v}_x)
    \label{eq:uxdot} \, , \\
  \frac{\dpa u_y}{\dpa t} + (\vc{u}\cdot\nab) u_y + u_y^{(0)} \frac{\dpa
    u_y}{\dpa y} &=& -\frac{1}{2} \varOmega_{\rm K} u_x \nonumber \\
    && \hspace{-4.0cm}-\frac{G M_{\rm p} y}{(x^2+y^2)^{3/2}}
    - c_{\rm s}^2 \frac{\dpa \ln
    \rho_{\rm g}}{\dpa y} - \frac{\epsilon}{\tau_{\rm f}} (u_y-\overline{v}_y)
  \label{eq:uydot} \, , \\
  \frac{\dpa\rho_{\rm g}}{\dpa t} + (\vc{u}\cdot\nab)\rho_{\rm g} + u_y^{(0)}   \frac{\dpa
    \rho_{\rm g}}{\dpa y} &=& -\rho_{\rm g} \nab \cdot \vc{u} \, .
\end{eqnarray}
In \Eq{eq:uxdot} we include the possibility of adding a radial pressure
gradient $\dpa\ln P/\dpa\ln r$ through the sub-Keplerian velocity
\begin{equation}
  \Delta v = - \frac{1}{2} \frac{H}{r} \frac{\dpa\ln P}{\dpa \ln r} c_{\rm s}\,
  .
\end{equation}
We use $\Delta v=0$ except in \S\ref{s:sandblast} where we set $\Delta v=0.05
c_{\rm s}$ \citep*{Nakagawa+etal1986}. The gas also feels the gravity of the
protoplanet. For the back-reaction friction force from the particles we
calculate the local solids-to-gas ratio through $\epsilon=\overline{\rho}_{\rm
p}/\rho_{\rm g}$.
\begin{figure*}
  \begin{center}
    \includegraphics[width=0.8\linewidth]{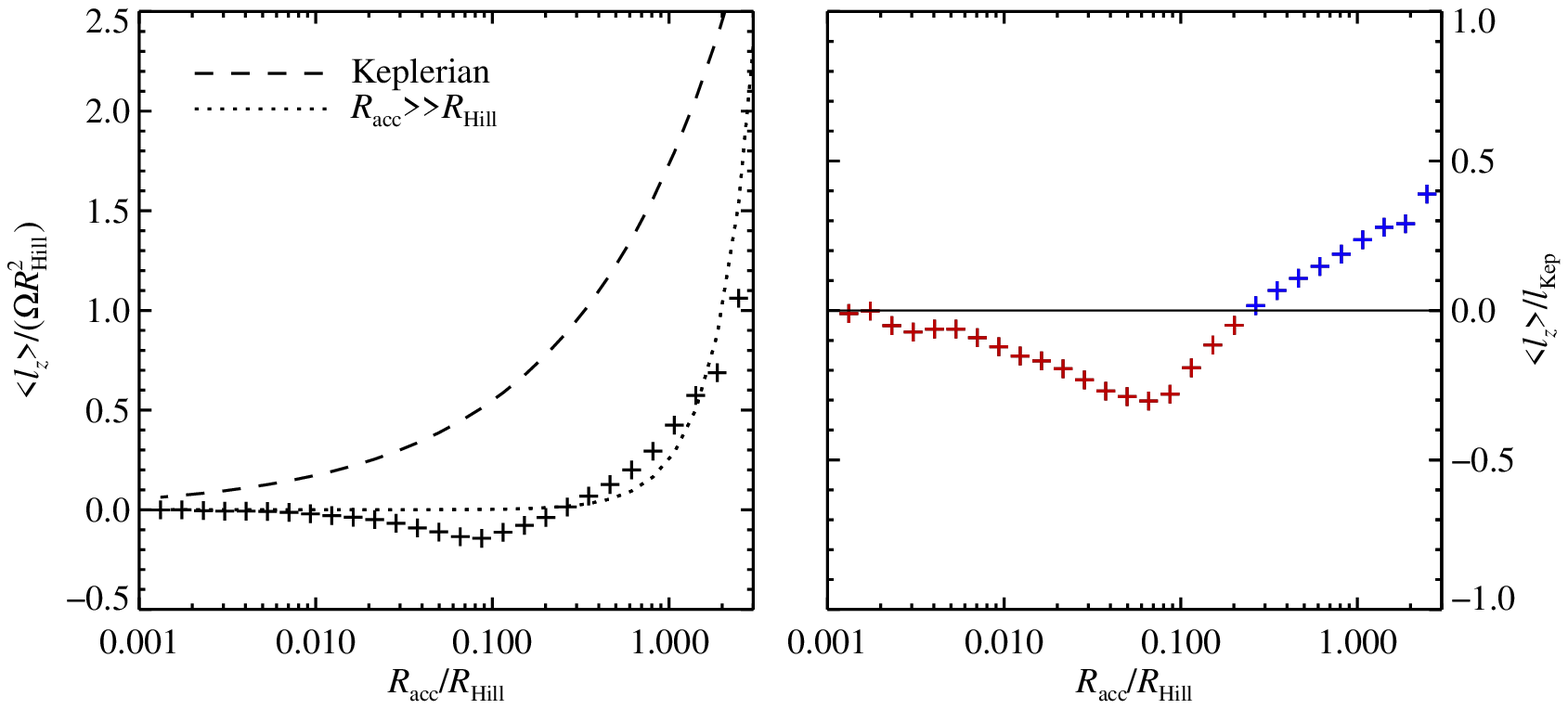}
  \end{center}
  \caption[]{Code validation test. The left plot shows the mean specific
  angular momentum, $\langle \ell_z \rangle$, of test particles passing circles
  of various radius $R_{\rm acc}$ ($x$-axis). In the limit $R_{\rm acc}\gg
  R_{\rm Hill}$ the slow prograde rotation of the nebula is transferred to the
  ``planet'' [the dotted line shows the analytical expectation
  $\ell_z=(1/4)\varOmega_{\rm K} R_{\rm acc}^2$], but particles rapidly
  decrease their angular momentum as they travel into the Hill sphere. The
  dashed line indicates prograde Keplerian specific angular momentum. The right
  plot shows the ratio of accreted angular momentum to the prograde Keplerian
  value. These non-interacting planetesimals induce a net retrograde rotation
  on the protoplanet. The measured angular momentum flux is in good agreement
  with results published in \cite{SchlichtingSari2007}.}
  \label{f:lz_racc}
\end{figure*}

The dynamical equations for the evolution of particles and gas appear in 3-D as
\begin{eqnarray}
  \dot{x}  &=& v_x \, , \\
  \dot{y}  &=& v_y - (3/2) \varOmega_{\rm K} x \, , \\
  \dot{z}  &=& v_z \, , \\
  \dot{v}_x &=& +2\varOmega_{\rm K} v_y-\frac{G M_{\rm p}
  x}{(x^2+y^2+z^2)^{3/2}} - \frac{1}{\tau_{\rm f}} (v_x - \overline{u}_x)
  \label{eq:xddot3D} \, , \\
  \dot{v}_y &=& -\frac{1}{2}\varOmega_{\rm K} v_x-\frac{G M_{\rm p}
  y}{(x^2+y^2+z^2)^{3/2}} - \frac{1}{\tau_{\rm f}} (v_y - \overline{u}_y)
  \label{eq:yddot3D} \, . \\
  \dot{v}_z &=& -\varOmega_{\rm K}^2 z-\frac{G M_{\rm p}
  z}{(x^2+y^2+z^2)^{3/2}} - \frac{1}{\tau_{\rm f}} (v_z - \overline{u}_z)
  \label{eq:yddot3D} \, ,
\end{eqnarray}
\begin{eqnarray}
  \frac{\dpa u_x}{\dpa t} + (\vc{u}\cdot\nab) u_x + u_y^{(0)} \frac{\dpa
    u_x}{\dpa y}&=& +2\varOmega_{\rm K} u_y \nonumber \\
    && \hspace{-4.0cm} - \frac{G M_{\rm p} x}{(x^2+y^2+z^2)^{3/2}} - c_{\rm s}^2
    \frac{\dpa \ln \rho_{\rm g}}{\dpa x} - \frac{\epsilon}{\tau_{\rm f}}
    (u_x-\overline{v}_x)
    \label{eq:uxdo3Dt} \, , \\
  \frac{\dpa u_y}{\dpa t} + (\vc{u}\cdot\nab) u_y + u_y^{(0)} \frac{\dpa
    u_y}{\dpa y} &=& -\frac{1}{2} \varOmega_{\rm K} u_x \nonumber \\
    && \hspace{-4.0cm}-\frac{G M_{\rm p} y}{(x^2+y^2+z^2)^{3/2}}
    - c_{\rm s}^2 \frac{\dpa \ln
    \rho_{\rm g}}{\dpa y} - \frac{\epsilon}{\tau_{\rm f}} (u_y-\overline{v}_y)
  \label{eq:uydo3Dt} \, , \\
  \frac{\dpa u_z}{\dpa t} + (\vc{u}\cdot\nab) u_z + u_y^{(0)} \frac{\dpa
    u_z}{\dpa y} &=& \nonumber \\
    && \hspace{-4.0cm}-\frac{G M_{\rm p} z}{(x^2+y^2+z^2)^{3/2}}
    - c_{\rm s}^2 \frac{\dpa \ln
    \rho_{\rm g}}{\dpa z} - \frac{\epsilon}{\tau_{\rm f}} (u_z-\overline{v}_z)
  \label{eq:uzdot3D} \, .
\end{eqnarray}
Here we subject particles to the vertical gravity component of the central
star, but ignore the same term for the pressure-supported gas.

\section{Code validation}
\label{s:validate}

To validate the implementation of the protoplanet gravity in the shearing sheet
frame, we test here the accretion of test particles against other published
results \citep{SchlichtingSari2007}.

We consider a protoplanet of dimensionless mass $GM_{\rm
p}=10^{-6}\varOmega_{\rm K}^2 H^3$ in a 2-D box of size $L_x=L_y=0.04 H$. Here
$\varOmega_{\rm K}$ is the Keplerian orbital frequency, while $H$ is the gas
scale height. The Hill radius of the protoplanet is $R_{\rm H}\approx0.007 H$.
We place 10,000 particles randomly in the box, initially excluding particles
from the region within the Hill sphere. The particle positions and velocities
evolve in response to the Coriolis force, the tidal force from the central star
and the gravity of the embedded protoplanet. For this test we ignore drag
forces between particles and gas.

As the particle positions and velocities evolve, we track the first crossing of
various distances from the protoplanet and calculate the mean specific angular
momentum per particle for all the accretion shells. The results are shown in
\Fig{f:lz_racc}. For accretion outside of the Hill sphere, the motion is
prograde, following the trend $\ell_z=(1/4)\varOmega_{\rm K} R_{\rm acc}^2$
derived analytically in \cite{DonesTremaine1993a}. However, deeper in the Hill
sphere the net angular momentum is approximately zero. Particles transported to
the vicinity of the protoplanet by the Keplerian shear lose their angular
momentum on the way and induce slow retrograde rotation to the protoplanet. The
numerical result shown in \Fig{f:lz_racc} is in good agreement with
\cite{SchlichtingSari2007}.

\section{Particle sizes}
\label{s:mmsn}

We characterize particles in terms of their dimensionless friction time
$\varOmega_{\rm K} \tau_{\rm f}$ (where $\varOmega_{\rm K}$ is the Keplerian
orbital frequency). The translation to a particle size depends on the assumed
disc model. Small particles are subject to Epstein drag, valid when the
particle radius is smaller than the mean free path of the gas (see below). The
friction time in the Epstein regime is given by
\begin{equation} 
  \varOmega_{\rm K} \tau_{\rm f}^{\rm (Ep)} = \frac{\varOmega_{\rm K}
  \rho_\bullet a}{\rho_{\rm g} c_{\rm s}} = \sqrt{2\pi} \frac{\rho_\bullet      a}{\varSigma_{\rm g}} \label{eq:tauEpstein}\, ,\\
\end{equation}
where the second step applies in the disc mid-plane. Here $\rho_\bullet$ is the
material density of the solids, $a$ is the radius of a solid body, $\rho_{\rm
g}$ is the gas density, $c_{\rm s}$ is the sound speed, while $\varSigma_{\rm
g}$ is the column density of gas. Epstein drag depends on gas density, but in
practice the gas density fluctuations are negligible (order 1\%) in our
subsonic flows, so we ignore them. Solving for particle size gives
\begin{eqnarray}
  a &=& \frac{\varOmega_{\rm K} \tau_{\rm f}^{\rm (Ep)} \varSigma_{\rm
  g}}{\sqrt{2\pi} \rho_\bullet}\nonumber \\&\approx& 60\,{\rm cm}\,\varOmega_{\rm K} \tau_{\rm f}^{\rm (Ep)}
  \left( \frac{\varSigma_{\mathrm g}}{300\,{\rm g\,cm^{-2}}} \right)
  \left( \frac{\rho_\bullet}{2\,{\rm g\,cm^{-3}}}
  \right)^{-1}\, .
\end{eqnarray}
Applying a simulation with given $\tau_{\rm f}$ values to different locations
in the disc with different gas column densities changes the physical particle
sizes corresponding to a given value of $\varOmega_{\rm K}\tau_{\rm f}$. We
follow the Minimum Mass Solar Nebula model (\citealt{Weidenschilling1977b};
\citealt{Hayashi1981}; \citealt*{Cuzzi+etal1993}), with column density
$\varSigma_{\rm g}$ and gas scale height $H$ given by
\begin{eqnarray}
  \varSigma_{\rm g}(r) &=& 1700\,{\rm g\,cm^{-2}}\,\left(\frac{r}{\rm
  AU}\right)^{-3/2} \, , \\
  H(r)         &=& 0.033\,{\rm AU}\,\left(\frac{r}{\rm AU}\right)^{5/4} \, .
\end{eqnarray}
Considering, as in the main text, $r=3\,{\rm AU}$ and $\varSigma_{\rm
g}=300\,{\rm g\,cm^{-2}}$ or $r=10\,{\rm AU}$ and $\varSigma_{\rm g}=50\,{\rm
g\,cm^{-2}}$, yields particle sizes of $a=6$ cm and $a=1$ cm, respectively, for
$\varOmega_{\rm K}\tau_{\rm f}=0.1$.
\renewcommand{\thefigure}{E1}
\begin{figure*}
  \begin{center}
    \includegraphics[width=0.32\linewidth]{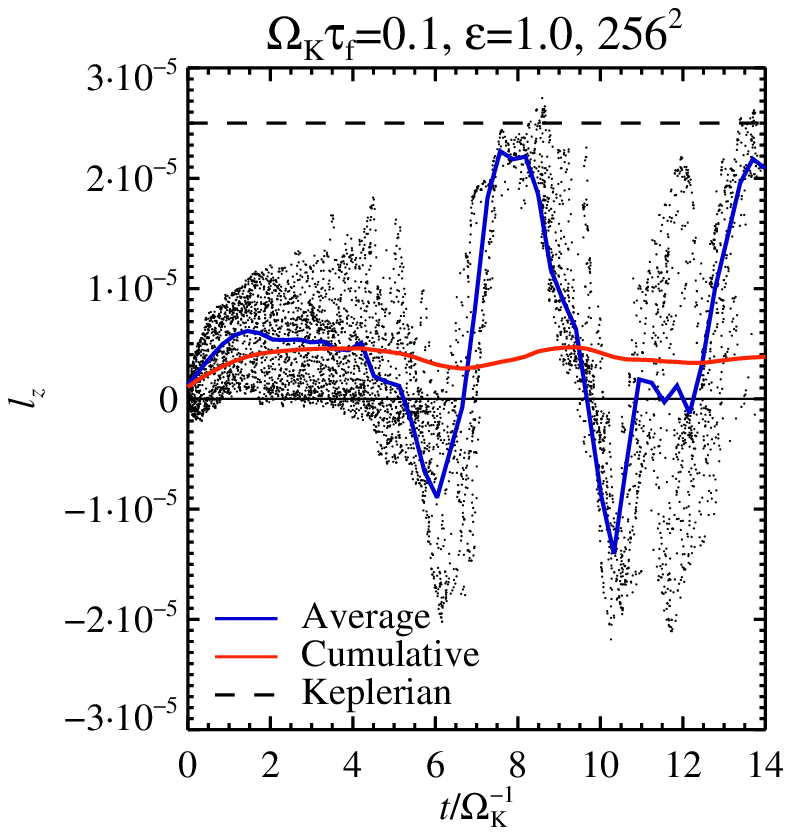}
    \includegraphics[width=0.32\linewidth]{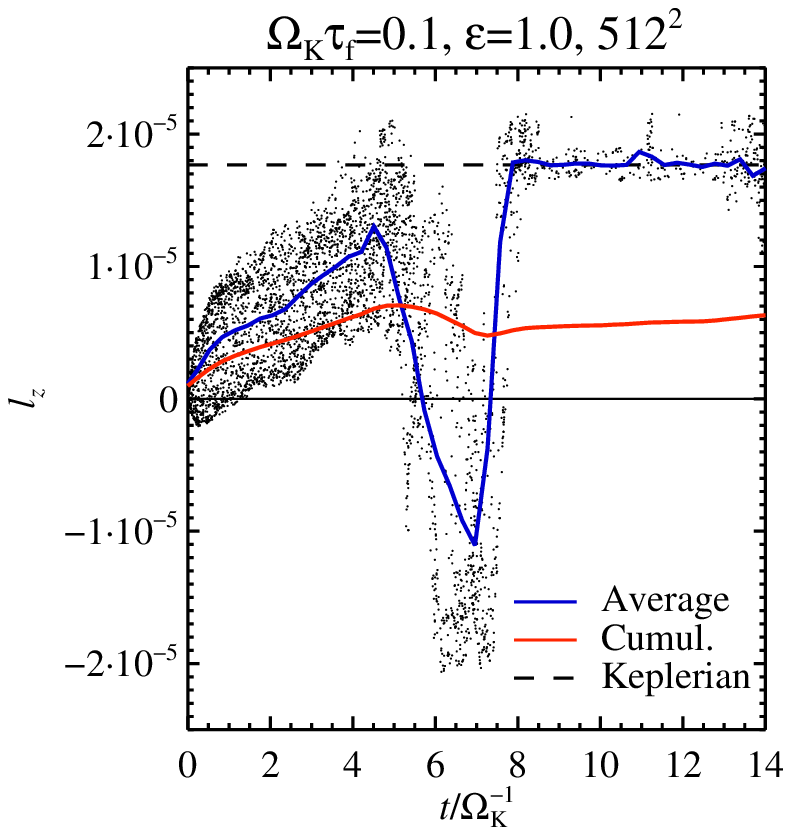}
    \includegraphics[width=0.32\linewidth]{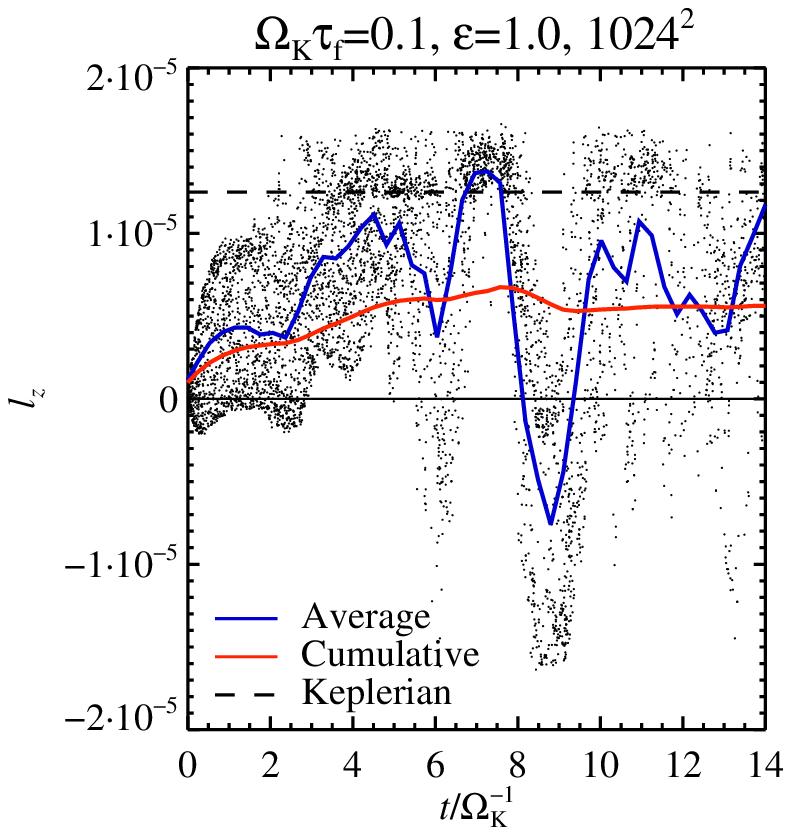}
  \end{center}
  \caption{Convergence in grid resolution. The plots show the accreted angular
  momentum for runs with $256^2$ grid points (left panel), $512^2$ grid points
  (middle panel) and $1024^2$ grid points (right panel). In all cases a net
  positive angular momentum is accreted, although periods of retrograde
  accretion occur as well. The inner radius of the simulations, where particles
  are considered to be accreted to the protoplanet, is reduced proportional to
  the resolution.}
  \label{f:lz_t_conv}
\end{figure*}
\renewcommand{\thefigure}{F1}
\begin{figure}
  \begin{center}
    \includegraphics[width=7.0cm]{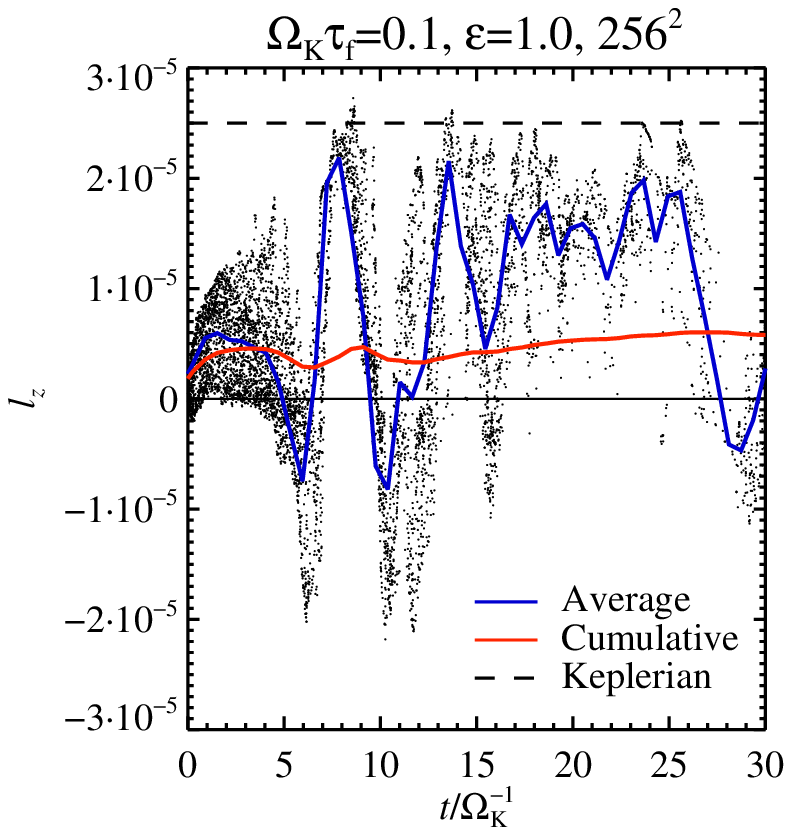}\\
    \includegraphics[width=7.0cm]{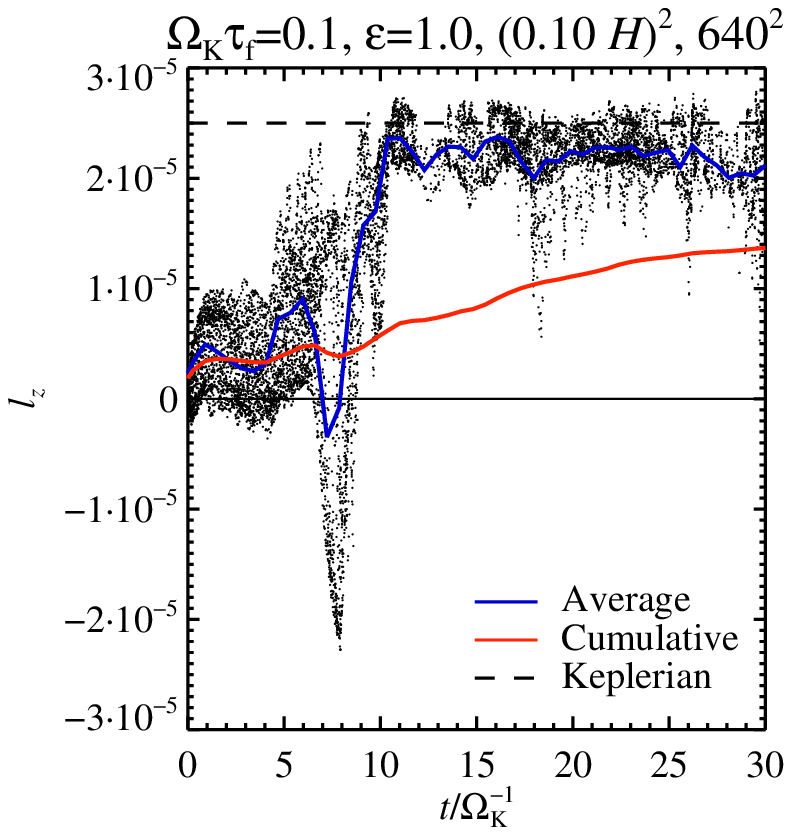}
  \end{center}
  \caption{Convergence in box size. The plots show the accreted angular
  momentum for low resolution runs with $L_x=L_y=0.04 H$ (top panel) and
  $L_x=L_y=0.1 H$ (bottom panel). Prograde motion is more readily accreted to
  the protoplanet in the larger box.}
  \label{f:lz_t_box}
\end{figure}

The Epstein regime of free molecular flow ceases to apply once the particle
radius exceeds (9/4 of) the gas mean free path,
\begin{eqnarray}
  \lambda &=& \frac{\mu}{\rho_{\rm g} \sigma_{\rm mol}} =
  \frac{\sqrt{2\pi}\mu H}{\varSigma_{\rm g} \sigma_{\rm mol}} \label{eq:mfp}\\
  &\approx & 31\, {\rm cm}
  \left(\frac{\varSigma_{\mathrm g}}{300\,{\rm g\,cm^{-2}}} \right)^{-1}
  \left( \frac{H/r}{0.043} \right)
  \left(\frac{r}{3 \rm AU}\right)\, ,
\end{eqnarray}
where $\mu = 3.9 \times 10^{-24}\,{\rm g}$ is the mean molecular weight and
$\sigma_{\rm mol} = 2 \times 10^{-15} {\rm cm}^2$ is the molecular cross
section of molecular hydrogen \citep{Nakagawa+etal1986}. Epstein drag applies
as long as
\begin{eqnarray} 
  \varOmega_{\rm K} \tau_{\rm f}^{\rm (Ep)} &<&  \frac {9 \pi}{2}
  \frac{\rho_\bullet \mu H}{\varSigma_{\rm g}^2 \sigma_{\rm mol}} \approx 1.2
  \left(\frac{\rho_\bullet}{2\,{\rm g\,cm^{-3}}}\right) \times \nonumber \\
  && \,\,\, \left(\frac{\varSigma_{\mathrm g}}{300\,{\rm g\,cm^{-2}}}
  \right)^{-2}\left(\frac{H/r}{0.043}\right)\left(\frac{r}{3\,{\rm
  AU}}\right)\, .
\end{eqnarray}
Thus Epstein drag is the relevant regime for the application of our model to
particles with $\varOmega_{\rm K}\tau_{\rm f}\lesssim1$ beyond $r=3\,{\rm AU}$.

\section{Particle collisions}
\label{s:collide}

In this section we argue that particle collisions are negligible compared to
gas drag in most of the simulation domain. The time-scale for collisions
between particles is
\begin{equation}
  \tau_{\rm c}=\frac{\lambda_{\rm p}}{c_{\rm p}} \, ,
\end{equation}
where $c_{\rm p}$ is the local velocity dispersion (``temperature'') of the
particles and $\lambda_{\rm p}$ is the mean free path for particle collisions.
The mean free path in a granular medium can be calculated from the particle
number density $n_{\rm p}$ and collisional cross section $\sigma_{\rm p}$,
\begin{equation}
  \lambda_{\rm p} = \frac{1}{n_{\rm p}\sigma_{\rm p}} \, .
\end{equation}
Assuming spherical particles of radius $a$, this yields a collision time-scale
\begin{equation}
  \tau_{\rm c}=\frac{a \rho_\bullet}{3 c_{\rm p}\rho_{\rm p}} \, .
\end{equation}
The ratio of Epstein friction time to collision time is (ignoring factors of
order unity)
\begin{equation}
  \frac{\tau_{\rm f}}{\tau_{\rm c}}\approx\frac{c_{\rm p}}{c_{\rm s}}
  \frac{\rho_{\rm p}}{\rho_{\rm g}} \, .
\end{equation}
Particle collisions become important when $\rho_{\rm p}/\rho_{\rm g}\gtrsim
c_{\rm s}/c_{\rm p}$, i.e.\ for flows with large particle loading. For the
typical value $c_{\rm p}\approx 0.01 c_{\rm s}$, particle collisions are
important when $\rho_{\rm p}/\rho_{\rm g}\gtrsim 100$. Thus we can ignore
particle collisions in the average state of the simulations presented in the
main paper. Nevertheless particle collisions will become important in the dense
regions close to the protoplanet, and indeed particle collisions may be
important for efficient accretion of circumplanetary material onto the
protoplanet.

\section{Convergence test}
\label{s:converge}

To quantify the effect of grid resolution and inner accretion radius on the
results presented in the main paper, we have run lower resolution ($256^2$ grid
points, 250,000 particles) and higher resolution ($1024^2$ grid points,
8,000,000 particles) variations of the main simulation. We set the inner
accretion radius to 4 grid points in all cases. The accreted angular momentum
is shown in \Fig{f:lz_t_conv}. It is clear that the behaviour at all three
resolutions is highly time variable, with periods of prograde and retrograde
accretion interspersed. However, in all cases a net positive angular momentum
is accreted, inducing between 25\% and 50\% of the surface Keplerian frequency
to the protoplanet. There is a trend for accreting stronger prograde rotation
with increasing resolution.

\section{Dependence on box size}
\label{s:boxsize}

We proceed to test the dependence of our results on the size of the simulation
domain. For this purpose we ran a low resolution simulation with the radial and
azimuthal extent expanded by a factor 2.5, to $L_x=L_y=0.1$. We kept the grid
and particle resolution constant by using $640\times640$ grid cells and
1,562,500 particles, corresponding in resolution to the $256^2$ simulation
shown in the left panel of \Fig{f:lz_t_conv}. The accreted angular momentum is
shown in \Fig{f:lz_t_box}. The larger box size has a net positive effect on the
prograde rotation of the protoplanet. The sign of the angular momentum is more
consistently positive than in the small box, and there are fewer periods of
retrogade accretion.

\bsp

\label{lastpage}


\begin{thebibliography}{99}

\bibitem[Adachi et~al.(1976)Adachi, Hayashi, \& Nakazawa]{Adachi+etal1976}
  Adachi I., Hayashi C., Nakazawa K., 1976, Progress of Theoretical
  Physics, 56, 1756
\bibitem[Alexander \& Armitage(2007)]{AlexanderArmitage2007}
  Alexander R.~D., Armitage P.~J., 2007, \mnras, 375, 500
\bibitem[Balbus \& Hawley(1998)]{BalbusHawley1998}
  Balbus S.~A., Hawley J.~F., 1998, Reviews of Modern Physics, 70, 1
\bibitem[Bizzarro et~al.(2005)Bizzarro, Baker, Haack, \&
Lundgaard]{Bizzarro+etal2005}
  Bizzarro M., Baker J.~A., Haack H., Lundgaard K.~L.,
  2005, \apjl, 632, L41
\bibitem[Brandenburg(2003)]{Brandenburg2003}
  Brandenburg A., 2003, Computational aspects of astrophysical MHD and
  turbulence (Advances in Nonlinear Dynamos), 269
\bibitem[Brauer et~al.(2008)Brauer, Henning, \& Dullemond]{Brauer+etal2008}
  Brauer F., Henning T., Dullemond C.~P., 2008, \aap, 487, L1
\bibitem[Brown(2008)]{Brown2008}
  Brown M.~E., 2008, The Largest Kuiper Belt Objects, ed. M.~A. Barucci,
  H.~Boehnhardt, D.~P. Cruikshank, \& A.~Morbidelli, 335
\bibitem[Brown et~al.(2007)Brown, Barkume, Ragozzine, \& Schaller]{Brown+etal2007}
  Brown M.~E., Barkume K.~M., Ragozzine D., Schaller E.~L., 2007,
  \nat, 446, 294
\bibitem[Canup(2005)]{Canup2005}
  Canup R.~M., 2005, Science, 307, 546
\bibitem[Chiang et~al.(2007)Chiang, Lithwick, Murray-Clay, Buie, Grundy, \& Holman]{Chiang+etal2007}
  Chiang E., Lithwick Y., Murray-Clay R., Buie M., Grundy W.,
  Holman M., 2007, in Protostars and Planets V, ed. B.~Reipurth,
  D.~Jewitt, K.~Keil, 895
\bibitem[Cuzzi et~al.(1993)Cuzzi, Dobrovolskis, \& Champney]{Cuzzi+etal1993}
  Cuzzi J.~N., Dobrovolskis A.~R., Champney J.~M., 1993, Icarus, 106, 102
\bibitem[Cuzzi et~al.(2008)Cuzzi, Hogan, \& Shariff]{Cuzzi+etal2008}
  Cuzzi J.~N., Hogan R.~C., Shariff K., 2008, \apj, 687, 1432
\bibitem[Davis \& Farinella(1997)]{DavisFarinella1997} Davis, D.~R., \&
  Farinella P., 1997, Icarus, 125, 50
\bibitem[Dominik et~al.(2007)Dominik, Blum, Cuzzi, \& Wurm]{Dominik+etal2007}
  Dominik C., Blum J., Cuzzi J.~N., Wurm G., 2007, in Protostars
  and Planets V, ed. B.~Reipurth, D.~Jewitt, K.~Keil, 783
\bibitem[Dones \& Tremaine(1993)]{DonesTremaine1993a}
  Dones L., Tremaine S., 1993, Icarus, 103, 67
\bibitem[Dones \& Tremaine(1993)]{DonesTremaine1993b}
  Dones L., Tremaine S., 1993, Science, 259, 350
\bibitem[Farinella et~al.(1992)Farinella, Davis, Paolicchi, Cellino, \&
Zappala]{Farinella+etal1992}
  Farinella P., Davis D.~R., Paolicchi P., Cellino A., Zappala V.,
  1992, \aap, 253, 604
\bibitem[Goldreich \& Tremaine(1980)]{GoldreichTremaine1980}
  Goldreich P., Tremaine S., 1980, \apj, 241, 425
\bibitem[Haghighipour \& Boss(2003)]{HaghighipourBoss2003}
  Haghighipour N., Boss A.~P., 2003, \apj, 583, 996
\bibitem[Hayashi(1981)]{Hayashi1981}
  Hayashi C., 1981, Prog.~Theor.~Phys., 70, 35
\bibitem[Jewitt(1999)]{Jewitt1999}
  Jewitt D., 1999, Annual Review of Earth and Planetary Sciences, 27, 287
\bibitem[Johansen et~al.(2007)Johansen, Oishi, Low, Klahr, Henning, \& Youdin]{Johansen+etal2007}
  Johansen A., Oishi J.~S., Mac Low M.-M., Klahr H., Henning T.,
  Youdin A., 2007, \nat, 448, 1022
\bibitem[Johansen et~al.(2009)Johansen, Youdin, \& Mac Low]{Johansen+etal2009}
  Johansen, A., Youdin, A., \& Mac Low, M. 2009, \apjl, 704, L75
\bibitem[Kato et~al.(2009)Kato, Nakamura, Tandokoro, Fujimoto, \& Ida]{Kato+etal2009}
  Kato M.~T., Nakamura K., Tandokoro R., Fujimoto M., Ida S.,
  2009, \apj, 691, 1697
\bibitem[Kokubo \& Ida(2007)]{KokuboIda2007}
  Kokubo E., Ida S., 2007, \apj, 671, 2082
\bibitem[Kretke \& Lin(2007)]{KretkeLin2007}
  Kretke K.~A., Lin D.~N.~C., 2007, \apjl, 664, L55
\bibitem[Kryszczy\'nska et~al.(2007)Kryszczy\'nska, La Spina, Paolicchi, Harris, Breiter, \& Pravec]{Kryszczynska+etal2007}
  Kryszczy\'nska A., La Spina A., Paolicchi P., Harris A.~W.,
  Breiter S., Pravec P., 2007, Icarus, 192, 223
\bibitem[Lacerda(2005)]{Lacerda2005}
  Lacerda P., 2005, PhD thesis, Leiden University, {\tt https://openaccess.leidenuniv.nl/handle/1887/603}
\bibitem[Lacerda \& Luu(2006)]{LacerdaLuu2006}
  Lacerda P., Luu J., 2006, \aj, 131, 2314
\bibitem[Laskar \& Robutel(1993)]{LaskarRobutel1993}
  Laskar J., Robutel P., 1993, \nat, 361, 608
\bibitem[Lissauer \& Kary(1991)]{LissauerKary1991}
  Lissauer J.~J., Kary D.~M., 1991, Icarus, 94, 126
\bibitem[Lommen et~al.(2009)Lommen, Maddison, Wright, van Dishoeck, Wilner, \&
Bourke]{Lommen+etal2009}
  Lommen, D., Maddison, S.~T., Wright, C.~M., van Dishoeck, E.~F.,
  Wilner, D.~J., \& Bourke, T.~L.
  2009, \aap, 495, 869
\bibitem[Lunine et~al.(2010)Lunine, O'Brien, Raymond, Morbidelli, Quinn, \&
Graps]{Lunine+etal2010}
  Lunine, J.~I., O'Brien, D.~P., Raymond, S.~N., Morbidelli, A., Quinn,
  T., \& Graps, A. 2010, Advanced Science Letters, in press (arXiv:0906.4369)
\bibitem[Machida et~al.(2008)Machida, Kokubo, Inutsuka, \& Matsumoto]{Machida+etal2008}
  Machida M.~N., Kokubo E., Inutsuka S.-i., Matsumoto T.,
  2008, \apj, 685, 1220
\bibitem[McAdoo \& Burns(1973)]{McAdooBurns1973}
  McAdoo D.~C., Burns J.~A., 1973, Icarus, 18, 285
\bibitem[Morbidelli et~al.(2009)Morbidelli, Bottke, Nesvorn\'y, \&
Levison]{Morbidelli+etal2009}
  Morbidelli, A., Bottke, W.~F., Nesvorn\'y, D., \& Levison, H.~F.
  2009, Icarus, 204, 558
\bibitem[Muto \& Inutsuka(2008)]{MutoInutsuka2008}
  Muto T., Inutsuka S.-i., 2009, \apj, 695, 1132
\bibitem[Nakagawa et~al.(1986)Nakagawa, Sekiya, \& Hayashi]{Nakagawa+etal1986}
  Nakagawa Y., Sekiya M., Hayashi C., 1986, Icarus, 67, 375
\bibitem[Rodmann et~al.(2006)Rodmann, Henning, Chandler, Mundy, \&
Wilner]{Rodmann+etal2006}
  Rodmann, J., Henning, T., Chandler, C.~J., Mundy, L.~G., \& Wilner, D.~J.
  2006, \aap, 446, 211
\bibitem[Safronov(1969)]{Safronov1969}
  Safronov V.~S., 1969, Evolution of the protoplanetary cloud and formation of
  Earth and the planets (Jerusalem: Israel Sci. Transl.)
\bibitem[Schlichting \& Sari(2007)]{SchlichtingSari2007}
  Schlichting H.~E., Sari R., 2007, \apj, 658, 593
\bibitem[Sheppard et~al.(2008)Sheppard, Lacerda, \& Ortiz]{Sheppard+etal2008}
  Sheppard S.~S., Lacerda P., Ortiz J.~L., 2008, Photometric
  Lightcurves of Transneptunian Objects and Centaurs: Rotations, Shapes, and
  Densities, ed. M.~A. Barucci, H.~Boehnhardt, D.~P. Cruikshank,
  A.~Morbidelli, 129
\bibitem[Sicilia-Aguilar et~al.(2007)Sicilia-Aguilar, Hartmann, Watson, Bohac,
Henning, \& Bouwman]{Sicilia-Aguilar+etal2007}
  Sicilia-Aguilar A., Hartmann L.~W., Watson D., Bohac C., Henning Th.,
  Bouwman J., 2007, \apj, 659, 1637
\bibitem[Testi et~al.(2003)Testi, Natta, Shepherd, \& Wilner]{Testi+etal2003}
  Testi L., Natta A., Shepherd D.~S., Wilner D.~J., 2003, \aap, 403, 323
\bibitem[Throop \& Bally(2005)]{ThroopBally2005}
  Throop H.~B., Bally J., 2005, \apjl, 623, L149
\bibitem[Weidenschilling(1977a)]{Weidenschilling1977}
  Weidenschilling S.~J., 1977a, \mnras, 180, 57
\bibitem[Weidenschilling(1977b)]{Weidenschilling1977b}
  Weidenschilling S.~J., 1977b, \apss, 51, 153
\bibitem[Weidenschilling \& Davis(1985)]{WeidenschillingDavis1985}
  Weidenschilling S.~J., Davis D.~R. 1985, Icarus, 62, 16
\bibitem[Wilner et~al.(2005)Wilner, D'Alessio, Calvet, Claussen, \& Hartmann]{Wilner+etal2005}
  Wilner D.~J., D'Alessio P., Calvet N., Claussen M.~J.,
  Hartmann L., 2005, \apjl, 626, L109
\bibitem[Yang et~al.(2007)Yang, Goldstein, \& Scott]{Yang+etal2007}
  Yang J., Goldstein J.~I., Scott E.~R.~D., 2007, \nat, 446, 888
\bibitem[Youdin \& Johansen(2007)]{YoudinJohansen2007}
  Youdin A., Johansen A., 2007, \apj, 662, 613

\end{thebibliography}
\end{document}